\documentclass{PoS}
\usepackage[latin9]{inputenc}
\usepackage{units}
\usepackage{amsmath}
\usepackage{subcaption}

\makeatletter
\title{The twisted gradient flow running coupling in SU(3): a non-perturbative determination}

\ShortTitle{The TGF non-perturbative running coupling in SU(3)}

\author{\speaker{Eduardo I. Bribián}%
\\
       Instituto de Física Teórica UAM-CSIC, Madrid\\
       Nicolás Cabrera 13-15, Universidad Autónoma de Madrid, E-28049-Madrid,Spain\\
       E-mail: \email{e.i.bribian@csic.es}}

\author{Margarita García Pérez%
\\
       Instituto de Física Teórica UAM-CSIC, Madrid\\
       Nicolás Cabrera 13-15, Universidad Autónoma de Madrid, E-28049-Madrid,Spain\\
       E-mail: \email{margarita.garcia@uam.es}}

\author{Alberto Ramos%
\\
       School of Mathematics and Hamilton Mathematics Institute, Dublin\\
       Trinity College Dublin, Dublin 2, Ireland\\
       E-mail: \email{alberto.ramos@maths.tcd.ie}}

\abstract{We report some preliminary results of our ongoing non-perturbative computation of the twisted 't Hooft running coupling in a particular set-up, using the gradient flow to define the coupling and step scaling techniques to compute it. For the computation we considered a pure gauge SU(3) theory in four dimensions, defined on the lattice on an asymmetrical torus endowed with twisted boundary conditions in a single plane, and related the energy scale of the coupling to an effective size combining the size of the torus and the rank of the gauge group. Additionally, we explore some of the effects of the freezing of the topology on the computation of the coupling.}

\FullConference{37th International Symposium on Lattice Field Theory - Lattice2019\\
		16-22 June 2019\\
		Wuhan, China}

\makeatother

\newcommand{\lt}{\tilde{l}}
\newcommand{\LT}{\tilde{L}}
\newcommand{\LTGF}{\lambda_{TGF}}
\newcommand{\MS}{{\overline{\rm MS}}}

\begin{document}

\section{Introduction}

	The gradient flow \cite{Narayanan:2006rf,Luscher:2009eq} has, in the last decade, become quite commonplace in the study of gauge theories. Particularly, and in combination with finite volume scaling methods, it has been widely used to explore the scale dependence of the coupling, with several renormalisation schemes being developed to this purpose~\cite{Luscher:2010iy,Fodor:2012td,Fritzsch:2013je,Ramos:2014kla}.

	We will in these proceedings report some preliminary results of our ongoing computation of the $SU(3)$ running coupling in the twisted gradient flow (TGF) scheme on the lattice. Our scheme is based on a modification to the finite volume scheme introduced in~\cite{Ramos:2014kla}, and was already used in~\cite{Bribian:2019} to compute the running coupling at NLO in continuum perturbation theory for several $SU(N)$ groups. Similar schemes have also been used in recent years in $SU(3)$ pure gauge theory to non-perturbatively determine the ratio of $\Lambda$ parameters with the $\MS$ scheme~\cite{Ishikawa:2017xam,Ramos:2019}, or, along with ideas of volume reduction~\cite{Eguchi:1982nm,GonzalezArroyo:1982ub,GonzalezArroyo:1982hz,GonzalezArroyo:1983ac,GonzalezArroyo:2010ss}, in the obtention of the $SU(\infty)$ running coupling on a single-site lattice, through the use of step scaling procedures based on rescaling the rank of the gauge group instead of the size of the lattice~\cite{GarciaPerez2015}, among other interesting computations~\cite{Perez:2014sqa,Keegan:2015lva,Ramos:2015,Perez:2017jyq}.

	The paper will be organised as follows: first, we will introduce the twisted gradient flow scheme, explaining our setup and our implementation of twisted boundary conditions. We will then detail how the gradient flow is used to define a renormalised 't Hooft coupling, and show the lattice implementation of the scheme and the step scaling procedure followed. Finally, we will detail how the simulations are being performed and present some preliminary results, showing along the way how some of the issues that popped up were addressed.

\section{The twisted gradient flow scheme}

	The first step towards obtaining the running of the coupling is to describe the scheme used to define it, which is identical to the one that was defined in~\cite{Bribian:2019}. 

	Let us consider a pure gauge $SU(3)$ theory, defined on an asymmetrical four-dimensional torus of sides $l$ in two directions and $\lt=3l$ in the other two. Twisted and periodic boundary conditions are respectively implemented for the shorter and longer sides, which translates into the following periodicity conditions for gauge fields:
	\begin{align}
			& A_\mu ( x + l \hat \nu) = \Gamma_\nu A_\mu (x) \Gamma_\nu^\dagger, &\text{ for } \nu=0,1, \nonumber \\
			& A_\mu ( x + \lt \hat \nu) = A_\mu (x), &\text{ for } \nu=1,2,
	\end{align}
	where $\Gamma_\nu$ denotes two $SU(3)$ matrices satisfying:
	\begin{equation}
		\Gamma_0 \Gamma_1  = Z_{10} \Gamma_1 \Gamma_0, \quad Z_{01} = Z^\star_{10} = \exp \{ 2 \pi i / 3 \} . \label{eq:twist}
	\end{equation}

	This choice of torus dimensions answers to symmetry reasons in the context of volume reduction, as this choice of boundary conditions forces momentum to be quantised in units of $2 \pi / \lt$ in all directions, as if we were dealing with a fully periodic torus of sides $\lt$. Moreover, such boundary conditions are incompatible with zero momentum modes, making it quite convenient if one wishes to implement perturbation theory.

	To define the coupling we used the gradient flow~\cite{Luscher:2010iy} in a similar manner to the one in~\cite{Ramos:2014kla}. As usual, one introduces the flow time parameter $t$ and defines new gauge fields $B_\mu(x,t)$ to be smeared towards the classical solutions of the e.o.m's by the flow equations:	
	\begin{equation}
		 \partial_{t}B_{\mu}(x,t)=D_{\nu}G_{\nu\mu}(x,t), \qquad  B_\mu(x,0)=A_\mu(x), \label{eq:floweqs}
	\end{equation}
	where $G_{\mu\nu}$ and $D_\mu$ denote the field strength tensor and covariant derivative of the flow fields:
		\begin{align}
			 &G_{\mu\nu}(x,t)=\partial_{\mu}B_{\nu}(x,t)-\partial_{\nu}B_{\mu}(x,t)+i\left[B_{\mu}(x,t),B_{\nu}(x,t)\right], \nonumber \\
			 &D_{\mu}B_{\nu}(x,t)=\partial_{\mu}B_{\nu}(x,t)+i\left[B_{\mu}(x,t),B_{\nu}(x,t)\right] .
		\end{align}

	It has been shown~\cite{Luscher:2010iy} that observables built from $B_\mu$ fields are renormalised quantities at $t>0$, allowing the following definition of a renormalised 't Hooft coupling~\cite{Ramos:2014kla}:
	\begin{equation}
		\LTGF(\tilde{l})= \left. \frac{t^2 }{N} \mathcal{F}(c) \left< E(t) \right> \right|_{t=\frac{1}{8}c^2\tilde{l}^2}, \quad E(t)= \frac{1}{2} \mathrm{Tr} (G_{\mu\nu}(x,t) G_{\mu\nu}(x,t) ) . \label{eq:observable}
	\end{equation}
	The factor $\mathcal{F}(c)$ is introduced so that $\LTGF=\lambda_0 + \mathcal{O}(\lambda_0^2)$. The running scale $\mu$ is related to both flow time and the effective size $\lt$ through $\mu^{-1}=\sqrt{8t}=c\lt$, where $c$ is a pre-chosen scheme-defining number. We explored the range $c \in [0.1,0.8]$ and found intermediate values to work better, as is well known from for instance~\cite{Fritzsch:2013je}.

\section{The TGF coupling on the lattice}

	Once the scheme is set up, the next step is to discretise the torus on a lattice of spacing $a$ and sides $l=La$ and $\lt = \LT a$. We replace the standard link variables to their flow versions $V_{\mu}(n,t)$, and then are left with three quantities to discretise: We need to pick the actions used to drive the Monte Carlo simulations, the flow equations in equation~\eqref{eq:floweqs}, and have to choose the observable used to define the coupling~\eqref{eq:observable}. The effects of these choices have been discussed in depth in~\cite{Ramos:2014kla,GarciaPerez2015,Ramos:2015}, so we merely mention our discretisation choices.

	For the first one, we used the Wilson plaquette action with twisted boundary conditions:
	\begin{equation}
		 S_w(V) = bN \sum_{n} \sum_{\mu\nu} \text{Tr }[ 1 - Z_{\mu\nu}(n)  P_{\mu\nu}(n,t) ], \quad P_{\mu\nu}(n,t)=V_\mu(n,t) V_\nu(n+\hat{\mu},t) V^\dagger_\mu(n+\hat{\nu},t) V^\dagger_\nu(n,t),
	\end{equation}
	where $Z_{\mu\nu}(n)$ is one for all plaquettes except the ones in the $01$ corner, which are as shown in~\eqref{eq:twist}. The same action was used in discretising the flow equations as well, though the full expression of the flow equation is a bit long for these proceedings. The explicit expressions can be found in~\cite{Ramos:2014kla}.

	As for the observable, two different discretisations were used to estimate the importance of artifacts, the plaquette and clover definitions of the energy density:
	\begin{equation}
		E_P(t) = \text{Tr }[ 1 - Z_{\mu\nu}(n)P_{\mu\nu}(n,t)], \qquad E_C(t) = \frac{1}{2} \text{Tr }[ \hat{G}_{\mu\nu}(n,t) \hat{G}_{\mu\nu}(n,t) ],
	\end{equation}
	where we defined, denoting $V_{-\mu}(n)=V^\dagger_\mu(n-\mu)$:
	\begin{align}
		\hat{G}_{\mu\nu}(n,t) \nonumber &= -\frac{i}{8} \{ Z_{\mu\nu}(n) P_{\mu\nu}(n,t) + Z_{\mu\nu}(n-\hat{\nu}) P_{-\nu\mu}(n,t)  \nonumber \\
		& + Z_{\mu\nu}(n-\hat{\mu}) P_{\nu-\mu}(n,t) + Z_{\mu\nu}(n-\hat{\mu}-\hat{\nu}) P_{-\mu-\nu}(n,t) -c.c. \} .
	\end{align}
	The $\mathcal{F}(c)$ factor in~\eqref{eq:observable} is substituted by a leading order computation on the lattice $\mathcal{F}(c,\LT)$, so that $\LTGF=\lambda_0 + \mathcal{O}(\lambda_0^2)$ for all lattice sizes.

	Step scaling was then used in the standard way to compute the running coupling: we defined a step scaling function $\sigma(u,s)$ determined by the change in the coupling when the energy is rescaled by a factor $s$, along with its lattice discretised version $\Sigma(u,s,1/\LT)$:
	\begin{equation}
		\Sigma(u,s,1/\LT)=\left.\lambda_{TGF}(s\LT,b) \right|_{u=\lambda_{TGF}(\LT,b)} \quad \underset{1/\LT \rightarrow  0}{\longrightarrow} \quad \sigma(u,s). \label{eq:stepscaling}
	\end{equation}
	We chose $s=2$. By measuring $\LTGF$ on a lattice $L^2 \times \LT^2$ and on twice the lattice sizes $(2L)^2\times (2\tilde L)^2$ at constant bare coupling $b$ we obtained estimates of the step scaling function eq. \eqref{eq:stepscaling}.

\section{Simulations and preliminary results}

	The simulations were performed using the same algorithm detailed in~\cite{Ramos:2014kla}, which uses a combination of a single heatbath and $\LT$ overrelaxations in each step, though adapted to the asymmetrical torus. We simulated five $L^2 \times \LT^2$ lattices, taking $\LT=12,18,24,36,48$, for bare couplings $b$ ranging from $0.35$ to $0.61$ and values of $c$ between $0.1$ to $0.8$. The simulations were started from a zero-action configuration followed by 2000 thermalisation steps, after which gradient flow measurements were taken after $\LT$ steps (to reduce autocorrelations), as integrating the flow equations is more expensive than Monte Carlo updates, and we used a small enough integrator to make integration errors negligible. Between 500 and 10000 measurements were collected for each lattice, the lower end corresponding to the more expensive larger lattices. More simulations are in progress to catch up to the smaller ones.

\begin{figure} [t]
		\begin{minipage}{0.48\linewidth}
			\includegraphics[page=2,width=1.00\linewidth]{Lambda_LO}
		\end{minipage}
		\begin{minipage}{0.48\linewidth}
			\includegraphics[width=1.00\linewidth]{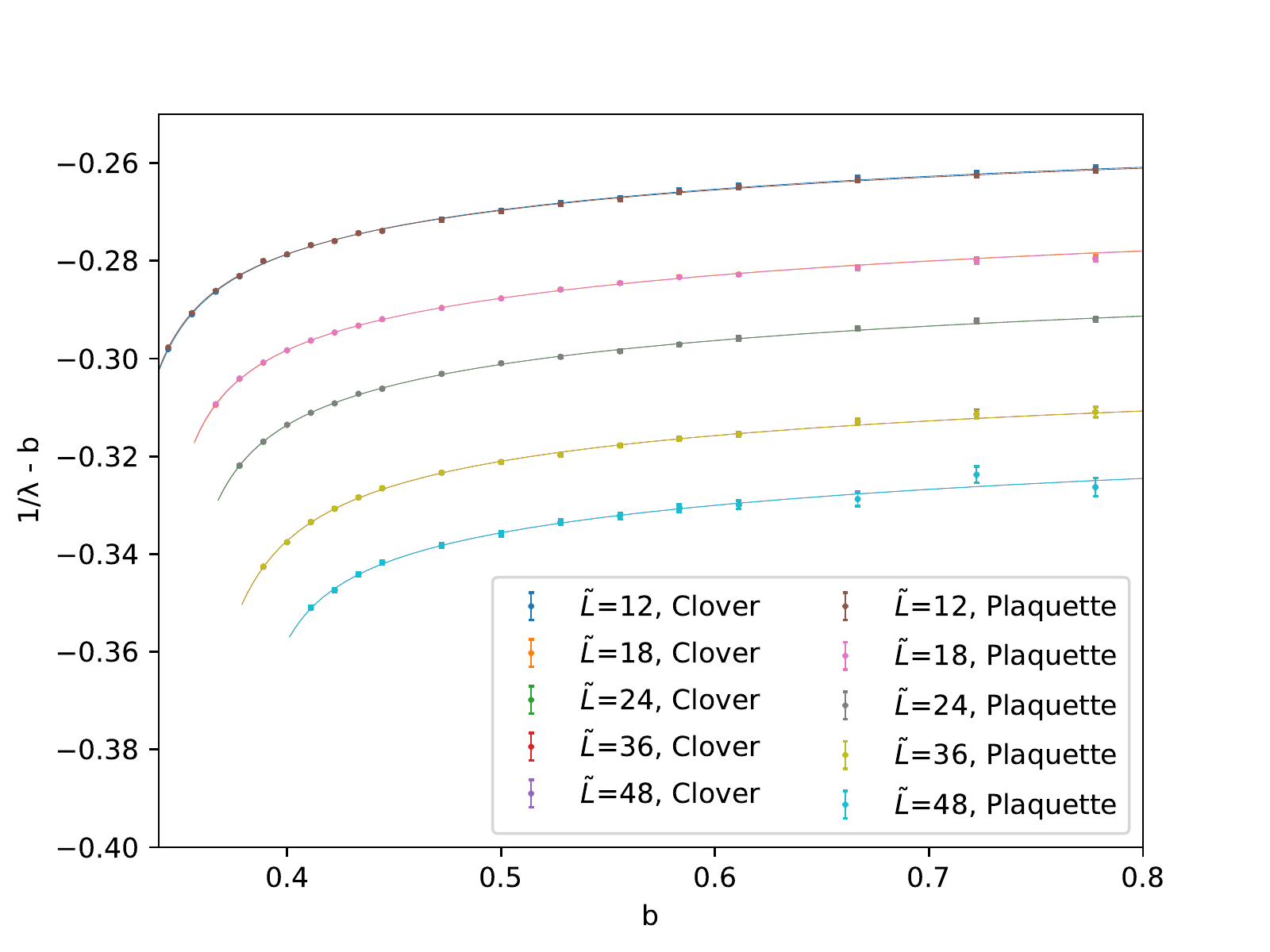}
		\end{minipage}
		\caption{Results for $b\LTGF$ at $c=0.3$ along with the corresponding NLO contribution, for the simulated values of $\LT$ and both the clover and plaquette observables (which are so close as to appear indistinguishable).}\label{fig:FullLambdaC030ocl}
		\centering
		\begin{minipage}{0.60\linewidth}
			\includegraphics[width=1.00\linewidth]{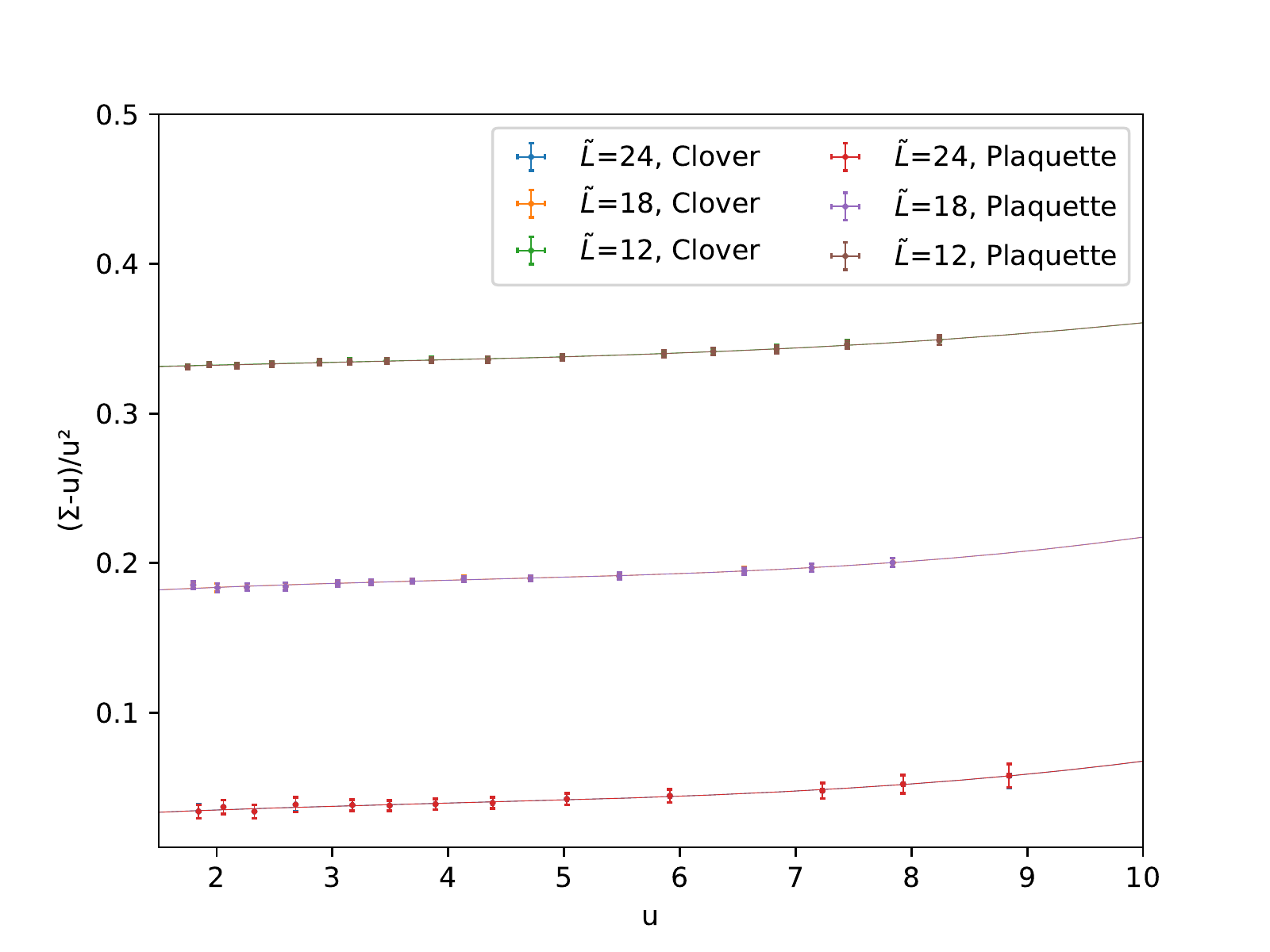}
		\end{minipage}
		%\caption{Results for the $\Sigma(u)/u$ functions at $c=0.3$, obtained from the $\LTGF$ values depicted above. The results have for clarity been displaced by 0.1 and 0.2 upwards for $\LT=18$ and $\LT=12$ respectively.}\label{fig:Full_Sigma_C30_ocl}
		\caption{Results for the $(\Sigma(u)-u)/u^2$ functions at $c=0.3$, obtained from the $\LTGF$ values depicted above. The results have for clarity been displaced by 0.15 and 0.30 upwards for $\LT=18$ and $\LT=12$ respectively.}\label{fig:Full_Sigma_C30_ocl}
	\end{figure}

	We will only show, in the scope of these proceedings, some illustrative examples of the preliminary results obtained for $c=0.3$. We show some results for both $\LTGF$ (fig.\ref{fig:FullLambdaC030ocl}) and for the step scaling functions (fig.\ref{fig:Full_Sigma_C30_ocl}), which are respectively fitted to a Padé-like Ansatz and a polynomial of the form:
	\begin{equation}
		\LTGF(\LT,b)=\frac{1}{b}\frac{a_0+a_1b + b^2}{a_2+a_3b+b^2}, \qquad \Sigma(u,\LT)= u + a_0 u^2 + a_1 u^3 + a_2 u^4 + a_3 u^5 ,
	\end{equation}
	which capture very well the behaviour of both quantities. The coupling is strongly dominated by its leading order term, and the next to leading order contribution is quite flat, meaning that intermediate values of $\LTGF$ obtained through interpolation should not incur in significant errors. The continuum extrapolation of the $\sigma$ function, however, is not yet obtained, as the statistics in the larger lattices are still being improved.
	
\begin{figure}[htbp]
		\centering
	  \begin{minipage}[b]{0.48\textwidth}
	    \includegraphics[width=\textwidth]{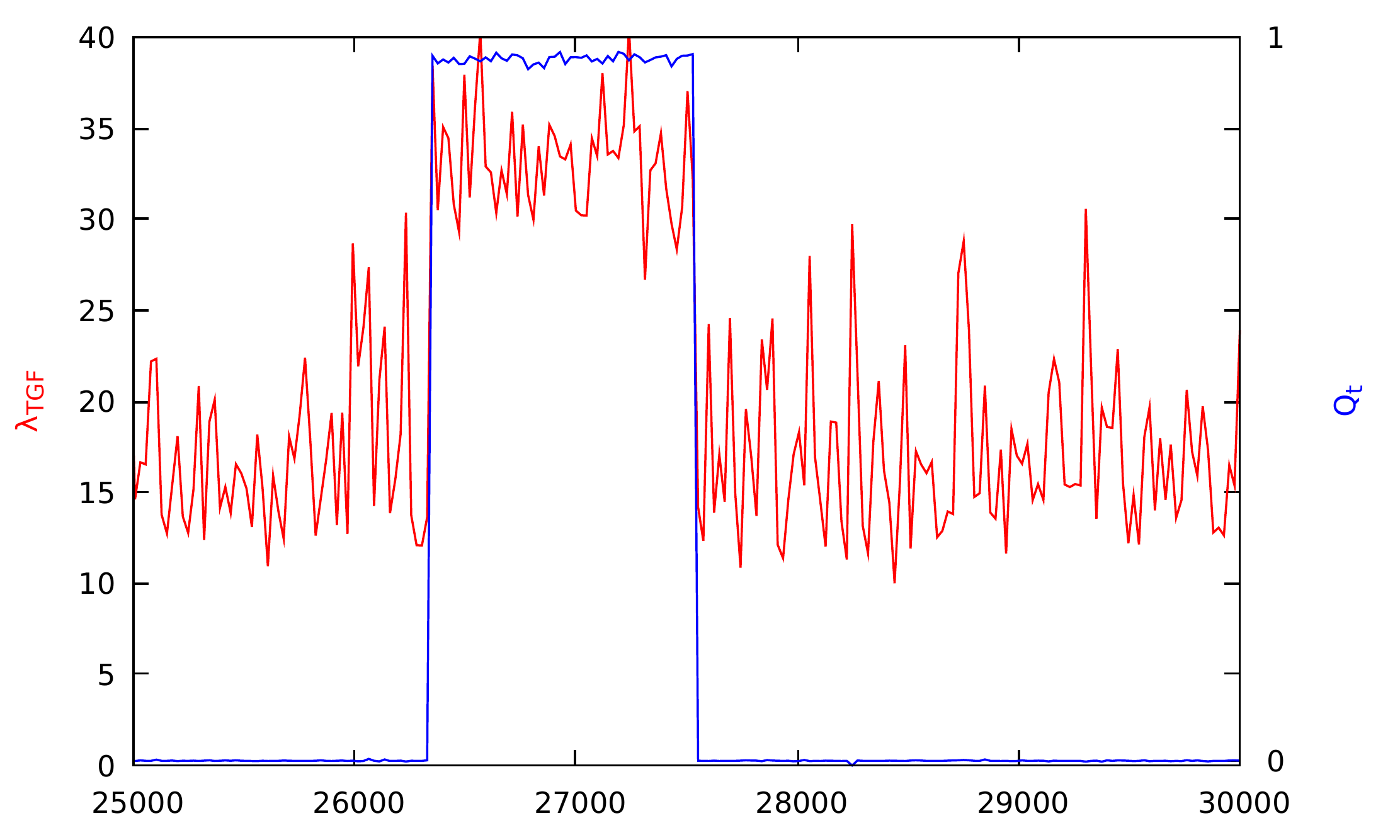}
	    \caption{An example of the strong correlation between the topological charge (blue) and the coupling $\LTGF$ (red) for the $\tilde{L}=24$ lattice at $b=0.38$ and with $c=0.3$.}
	    \label{fig:Freezing_L24B75}
	  \end{minipage} \quad
	  \begin{minipage}[b]{0.48\textwidth}
	    \includegraphics[width=\textwidth]{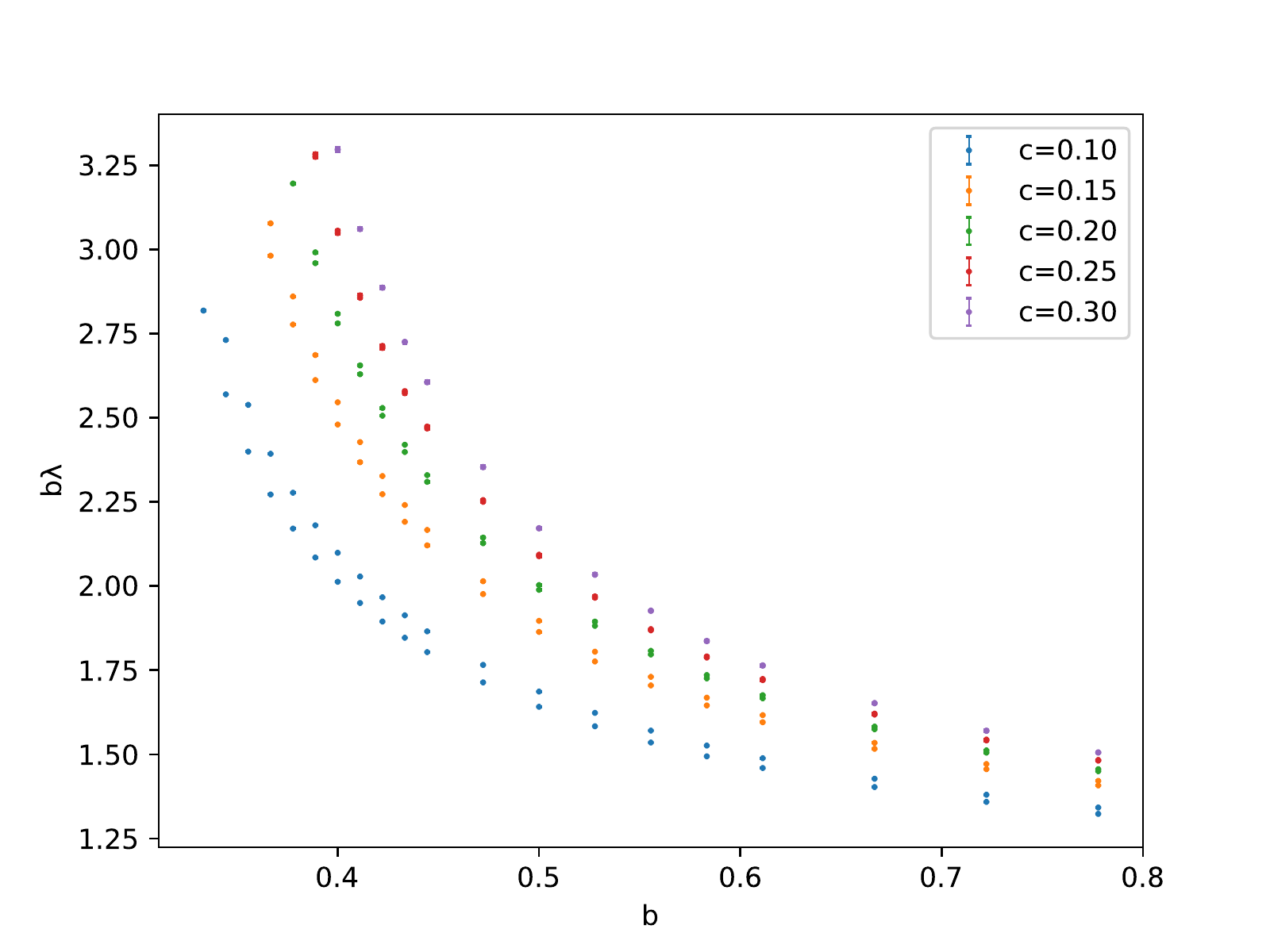}
	    \caption{Results for $b\lambda$ for $\LT=12$ with both observables shown in the same colour, to illustrate how lattice artifacts are quickly suppressed as $c$ increases.}
	    \label{fig:Issues_obs}
	  \end{minipage}
	  \bigskip 
		\begin{minipage}{0.48\linewidth}
			\includegraphics[page=1,width=1.00\linewidth]{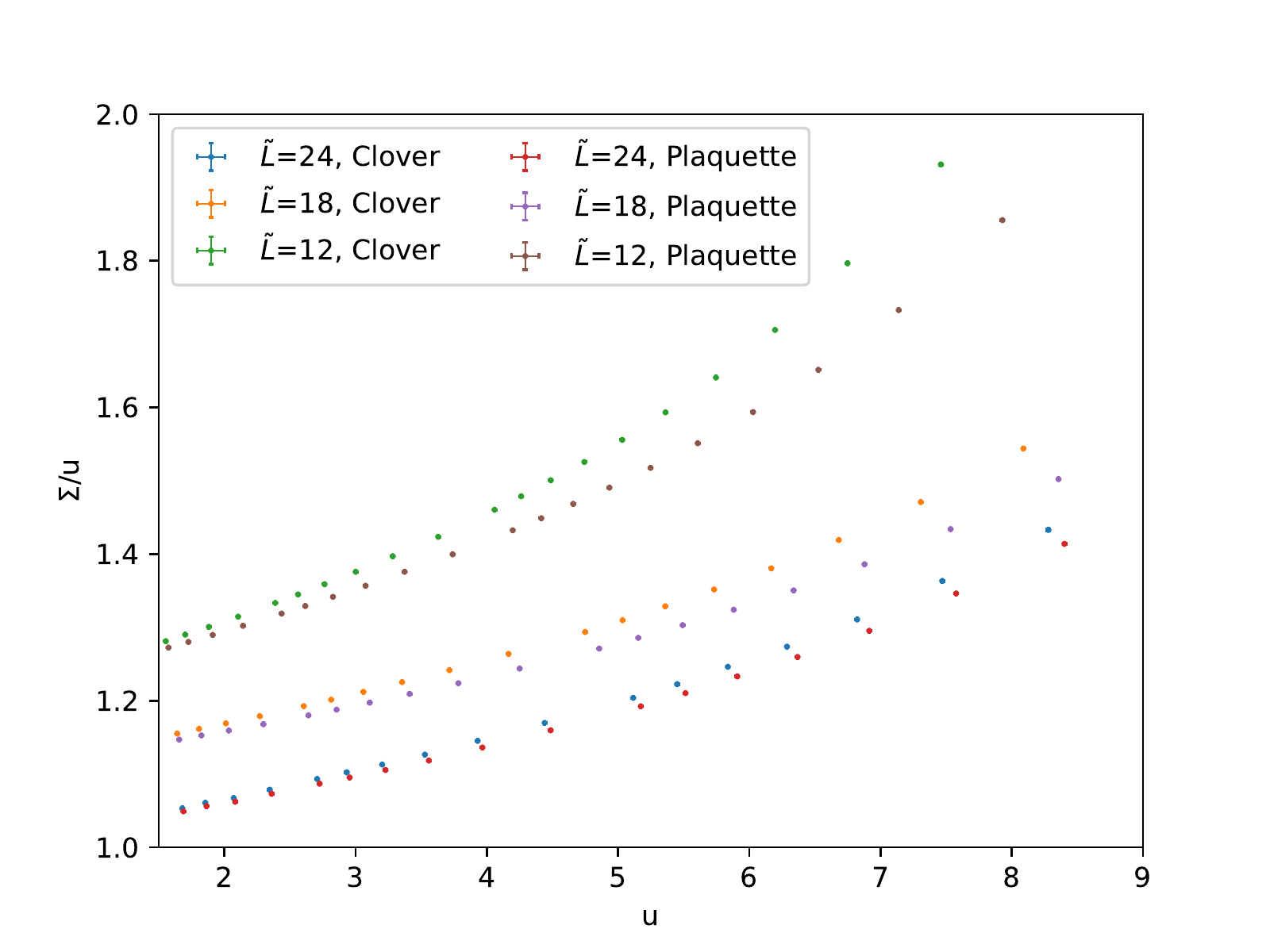}
		\end{minipage}
		\begin{minipage}{0.48\linewidth}
			\includegraphics[page=3,width=1.00\linewidth]{Sigma_LO}
		\end{minipage}
		\caption{Results for $\Sigma(u)/u$ for the clover observable at $c=0.10$ (left) and $c=0.70$ (right), exhibiting the effect of lattice artifacts at small $c$ and of the worsening of statistics at large $c$.}\label{fig:Issues_c}
	\end{figure}

	A few issues, however, had to be addressed. Namely, there were some complications related to topological freezing, and the effect of the flow (controlled by $c$) had some consequences in both statistics and artifacts. 	
	For the former, our observable turned out to be particularly sensitive to topology (though any observable should be plagued by similar effects), see fig.\ref{fig:Freezing_L24B75} for an example. Though our simulations were not completely frozen, a regime (corresponding to intermediate values of $\lt$ and small lattice spacing) appeared in which configurations tended to stall and stay in a given sector for a long time, leading to very large autocorrelations. 

	These large autocorrelations substantially increase the errors, and make it very difficult to estimate the uncertainties without access to ridiculously large statistics. We followed the proposal of \cite{Fritzsch:2013yxa} to circumvent the problem, modifying eq. \eqref{eq:observable} to:
	\begin{equation}
		\LTGF = t^2 \mathcal F(c, L) \frac{\langle E(t) \hat \delta(Q) \rangle }{\langle \hat \delta(Q) \rangle}\,,
	\end{equation}
	where $Q$ is the topological charge measured on the lattice (using the flow field) and $\hat \delta(Q)$ is zero for $|Q|>0.5$. This allows us to determine $\sigma(u,s)$ even in the region of parameters severely affected by topological freezing.

	%This situation had already been observed by the $\alpha$-Collaboration, which addressed it in a simple way: as a part of the scheme prescription, every single measurement for which the topological charge was nonzero is discarded. While this makes the simulations more expensive, we chose to adopt the same approach.

	The effects of the flow are similar to the ones observed in ~\cite{Ramos:2014kla}. Shorter flows (small $c$) lead to larger artifacts, whereas longer ones yield worse statistics. This completely overshadows the effect of the choice of discretised observable, to the point where beyond $c=0.25$ both observables yield perfectly compatible results. In the end, intermediate values of $c$, such as $c=0.3$, are a good compromise, combining smaller artifacts and good statistics. Figs.\ref{fig:Issues_c} and \ref{fig:Issues_obs} illustrate these issues.

\section{Summary and future prospects}

	We have used the gradient flow to define a renormalised 't Hooft running coupling for an $SU(3)$ theory on a four-dimensional, asymmetrical torus with twisted boundary conditions in one plane and periodical ones in the rest. Using the effective size of the torus as the running scale, we discretised the theory on the lattice, and are in the midst of computing the running coupling via step scaling, using lattices of sizes $\LT=12,18,24,36,48$.

	Preliminary results are in line with what was expected from previous similar step scaling studies, though we still need to extrapolate the results to the continuum, once the full simulations are finished, for a proper comparison. We plan to extend this analysis to other $SU(N)$ groups to explore the extent of finite $N$ effects in the context of volume reduction.

\section*{Acknowledgements}

	We would like to thank Antonio Gonz\'alez-Arroyo for many valuable discussions on both this topic and related ones. We acknowledge financial support from the MINECO/FEDER grant FPA2015-68541-P, the MINECO Centro de Excelencia Severo Ochoa Program SEV-2016-0597 and the EU grant H2020-MSCA-ITN-2018-813942 (EuroPLEx). E.I. Bribi\'an acknowledges support under the FPI grant BES-2015-071791. The numerical computations have been carried out at the IFT Hydra cluster and with computer resources provided by CESGA (Galicia Supercomputing Centre).

\providecommand{\href}[2]{#2}\begingroup\raggedright\endgroup

\end{document}